\documentclass[aip,apl,10pt, twocolumn]{revtex4}
\usepackage[dvipdfm]{graphicx,color}
\usepackage{dcolumn}
\usepackage{bm}
\usepackage{amsmath,amsthm,amssymb,mathrsfs}
\usepackage{subfigure}
 
\newcommand{\kv}{{\bm k}} 

\newcommand{\lt}{\left}
\newcommand{\rt}{\right} 
\newcommand{\qv}{{\bm q}}
\newcommand{\ve}{\varepsilon}

\newcommand{\kf}{k_{\rm F}} 
\newcommand{\e}{\epsilon}
 
\newcommand{\Rv}{\bm{R}} 
\newcommand{\ef}{\epsilon_{\rm F}} 
\begin{document}
\title{Dielectric Environment Effect on Carrier Mobility 
of Graphene Double-Layer Structure}

\author{Kazuhiro Hosono and Katsunori Wakabayashi}
\affiliation{
International Center for Materials Nanoarchitectonics (WPI-MANA), 
National Institute for Materials Science (NIMS), Namiki 1-1, Tsukuba 305-0044, Japan
}
\date{\today}
\begin{abstract}
 We have theoretically studied the dielectric environment effect on
 the charged-impurity-limited carrier mobility of graphene double-layer
 structure (GDLS) on the basis of 
 the Boltzmann transport theory. In this system, two graphene layers are
 separated by a dielectric barrier layer. It is pointed out that the 
 carrier mobility strongly depends on the dielectric constant of the
 barrier layer when the interlayer distance becomes larger than the
 inverse of the Fermi wave vector. Moreover, the conditions to improve the
 charged-impurity-limited carrier 
 mobility of the GDLS are evaluated.  
\end{abstract}

\maketitle

\begin{figure}
\includegraphics[width=0.48\textwidth]{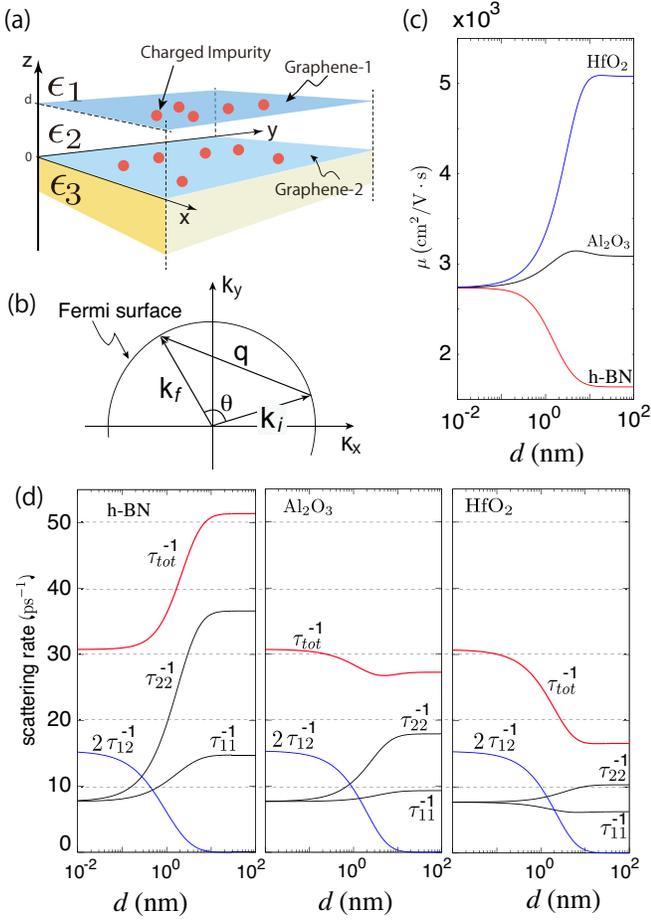}
 \caption{
(a) Schematic of graphene double-layer structure (GDLS)
 with three different dielectrics. The interlayer distance between
 two graphene layers is defined as $d$. Dielectrics with $\e_1$ and $\e_3$
 are assumed to be much thicker than the interlayer distance $d$. The top and
 bottom graphene layers are numbered $1$ and $2$. The red circles
 represent randomly distributed charged impurities.
(b) The carrier scattering process on a Fermi surface of radius $k_F$. An initial
state with a wavevector $\bm{k_i}$ is scattered by a charged impurity potential to
 a final state with a wavevector $\bm{k_f}$, where
 $|\bm{k_i}|=|\bm{k_f}|=k_F$. 
  Here
 $\bm{q}=\bm{k_f}-\bm{k_i}$, and $\theta$ is the scattering angle. 
(c) Interlayer distance $d$ dependence of the mobility for
three different inner barrier dielectrics, that is, 
 $h$-BN, Al$_2$O$_3$ and HfO$_2$, whose dielectric constants are
 $\e_{h-{\rm BN}}=4$, $\e_{\rm Al_2O_3}=12.53$ and $\e_{\rm HfO_2}=22$, respectively. 
$\e_1=\e_{\rm Air}=1$ and $\e_3=\e_{\rm Al_2O_3}$ are chosen. 
Here the carrier concentration $n_c$ and impurity concentration 
$n_{\rm i}$ are chosen as $n_c=10^{12}/{\rm cm^2}$, and 
$n_{\rm i}={5 \times }10^{11}/{\rm cm^2}$, respectively. 
(d) Scattering rate due to charged-impurity for
 three different inner layers. Left, middle and right panels represent
 the cases of $h$-BN, Al$_2$O$_3$ and HfO$_2$, respectively.
} 
\label{fig:1}
\end{figure}

Recent progress in graphene research has stimulated
the fabrication of new functional electronic devices which are composed of 
graphene and atomically-thin materials.
One such superlattice system is a graphene double-layer structure (GDLS), 
in which two graphene layers are separated by a thin dielectric, as shown in Fig.~\ref{fig:1}(a). 
Theoretically the GDLS is considered to be a good platform for studying the exciton
superfluidity~\cite{Pikalov2012, Mink2012}, the Coulomb drag
effect~\cite{Mink2012, Scharf2012} and plasmon mode~\cite{Profumo2012,
Badalyan2012, Stauber2012}. 
Recently, an optical device using this system was also proposed as an application~\cite{Novoselov2012}. 
In experiments, Al$_2$O$_3$\cite{Kim2011b} or $h$-BN
\cite{Dean2010,Ponomarenko2011,Lee2012, Zomer2012,Gorbachev2012} is used as the inner barrier layer. 
A recent experiment demonstrated that 
two graphene layers can be electrically separated by 
inserting a few atomic layers of $h$-BN, which correspond to the interlayer distance
$d \simeq 1$ nm~\cite{Gorbachev2012}. 

Carrier mobility is one of the key benchmarks of device performance
because it determines the power dissipation and switching speed. 
Recent theory suggests improving
carrier mobility by placing a high-$\kappa$ overlayer on the semiconductor nanostructure, 
which leads to weakening of Coulomb scattering due to the screening effect~\cite{Jena2007,Adam2007}.
Indeed, several electronic transport measurements on graphene or atomically-thin material have successfully revealed mobility enhancement
though change in the dielectric environment~\cite{Jang2008a, Kim2009, Radisavljevic2011,Hollander2011}.
However, a systematic study of the 
dielectric environment effect on the carrier mobility 
of the GDLS has not been performed yet. 

In this letter, we evaluate 
the dielectric environment effect on the charged-impurity-limited
carrier mobility of the GDLS
on the basis of the Boltzmann transport theory. 
We consider the static screening of Coulomb interaction within random
phase approximation (RPA). 
It is found that
carrier mobility strongly depends on the dielectric constant of the
barrier layer if the interlayer distance becomes larger than the
inverse of the Fermi wave vector. 
Our results offer effective use of ultra-thin dielectric barriers 
and a practical design strategy to improve the charged-impurity-limited
mobility of the GDLS.

Figure~\ref{fig:1}(a) shows a schematic of the GDLS, in which
two graphene layers are separated by three different
dielectrics $\e_1,\e_2$ and $\e_3$.
We assume that the two graphene layers are coupled only through 
Coulomb interaction between the charged impurities and carriers. The
wavefunction of the carrier in a graphene layer 
is described as the $\delta$ function in the $z$-direction~\cite{Ando2006}.
The Hamiltonian can be written as
\begin{align}
H=&\gamma\sum_{\kv, s, s^\prime}\sum_{i=1}^2c^{\dagger}_{\kv,s,i}\lt(\sigma_x k_x
 +\sigma_y k_y\rt)c_{\kv,s^\prime,i} 
\\
+&\frac{1}{L^2}\sum_{\kv\qv}\sum_{s,s'}\sum_{i,j} W_{ij}(q,
 d)c^{\dagger}_{\kv+\qv, s, i}c_{\kv, s', i}\rho^{(j)}_{\rm imp}(\qv), \nonumber 
\end{align}
where $c^{\dagger}_{\kv,s,i}$ ($c_{\kv,s,i}$) is the creation
(annihilation) operator for an electron with wavevector $\bm{k}=$($k_x$,$k_y$) and
pseudospin $s$ on the $i$-th graphene layer. Here $\gamma = 6.46$
eV$\cdot$\AA\ is the band parameter. $\sigma_x$ and $\sigma_y$ are the
Pauli spin matrices for pseudospin; $s,s'=\pm1$ are pseudospin labels to
describe the sublattice of the honeycomb lattice. 
$L^2$ is the area of each graphene layer.  $W_{ij}(q,d)$ denotes the Fourier
component of the screened Coulomb potential, which depends on the interlayer
distance $d$, and includes the effect of Coulomb interaction
between carriers on each graphene layer through the polarization
function~\cite{Ando2006}. 
$\rho^{(j)}_{\rm imp}(\qv)=\sum^{N_{imp}}_{\alpha}e^{-i\qv\cdot \Rv^{(j)}_{\alpha}}$ is the
particle density of random impurities on the $j$-th graphene layer having
total number of impurities $N_{imp}$. $\Rv^{(j)}_\alpha$ represents
the position of the impurities on the $j$-th layer.   
 \begin{figure*}[hbt]
\includegraphics[width=0.9\textwidth]{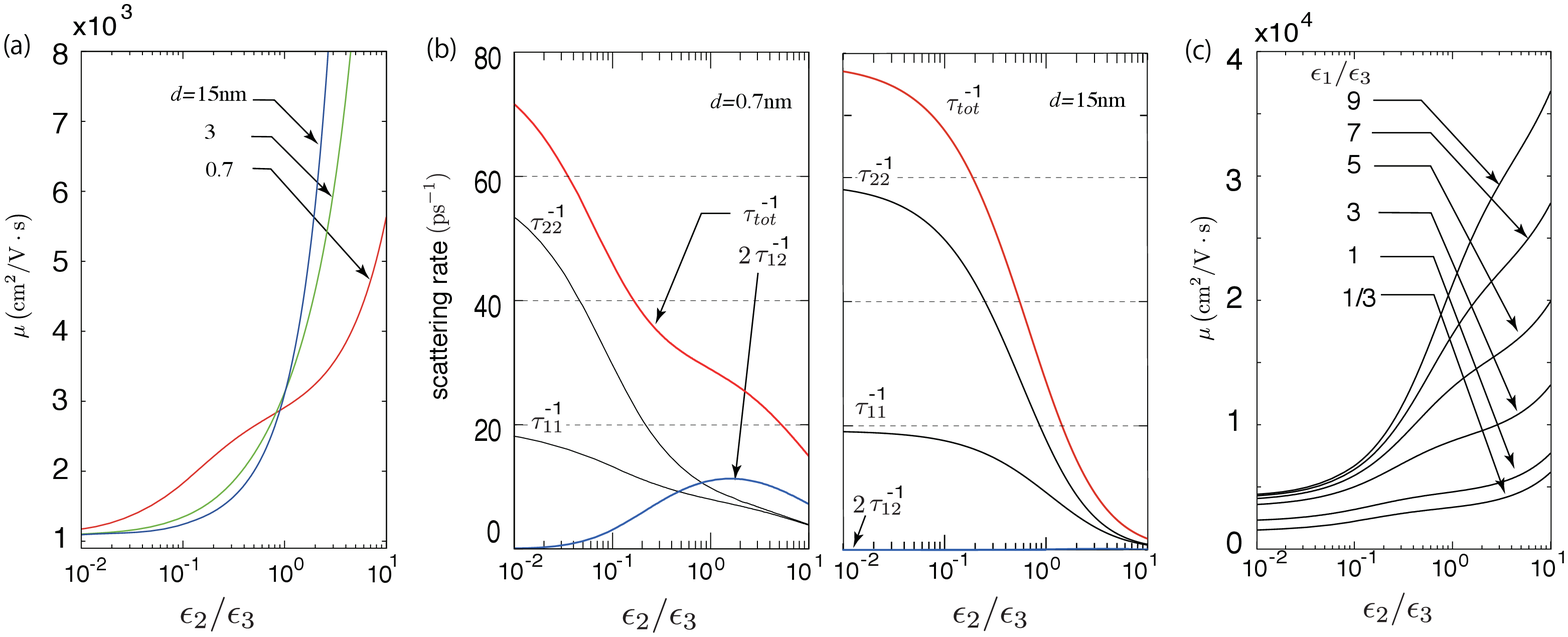}
 \caption{(a) Mobility versus inner dielectric constant for different
  interlayer distance, $d=0.7$, $3$ and $15$ nm. 
  Here the carrier concentration is $n_c=10^{12} /{\rm cm^2}$, the impurity
  concentration is $n_{\rm i}=5 \times 10^{11}/{\rm cm^2}$, 
  $\e_1=\e_{\rm Air}=1$ and $\e_3=\e_{\rm Al_2O_3}=12.53$. 
(b) The scattering rate as function of the barrier
  layer dielectric constant $\e_2$ for $d=0.7$ (left) and $15$ nm (right).
(c) The carrier mobility versus inner dielectric constant for various
  ratios of the top dielectric constant $\e_1$ to the bottom one $\e_3$ at
  $d=0.7$ nm and $\e_3=\e_{\rm Al_2O_3}=12.53$. }  
 \label{fig2}
 \end{figure*}

The carrier mobility $\mu$ can be described by $\mu=\frac{\sigma}{n_c e}$
using the carrier concentration $n_c$ and conductivity $\sigma$;
$\sigma=e^2D(\ef)\frac{\gamma^2}{2\hbar^2}\lt< \tau(\kf) \rt>$
\cite{Nomura2006,Ando2006a}. Here we assume that 
the imbalance of carrier concentration between two graphene layers
is absent. The total scattering rate at Fermi
energy is given as 
${\lt<\tau (\kf)\rt>}^{-1}\equiv
{\tau^{-1}_{tot}}={\tau^{-1}_{11}}+{\tau^{-1}_{22}}+{\tau^{-1}_{12}}+{\tau^{-1}_{21}}$. 
Here $\tau^{-1}_{11}$ ($\tau^{-1}_{22}$) is the intralayer scattering
rate of the first (second) graphene layer, and
$\tau^{-1}_{12}$ and $\tau^{-1}_{21}$ are interlayer scattering rates. 
According to the semiclassical Boltzmann theory, the momentum scattering rate is given by
\begin{align}
\frac{1}{\tau_{ij}(k)}=&n^{\rm imp}_{ij} \frac{D(\ve)}{\hbar}\int^{\pi}_{0}d\theta \lt|W_{ij}\lt(q, d\rt)\rt|^2(1-\cos^2\theta),
\end{align}
where $n^{\rm imp}_{11} (n^{\rm imp}_{22})$ is the impurity
concentration on first (second) graphene layer. The impurity
concentration for interlayer scattering is given as the average of
two layers $n^{\rm imp}_{12}\equiv \frac{n^{\rm imp}_{11}+n^{\rm
imp}_{22}}{2}$. 
For simplicity, we assume that the impurity concentration at each layer
$n_{\rm i}=n^{\rm imp}_{11}=n^{\rm imp}_{22}$ are equivalent
and that the Fermi level of both graphene layers lies in the conduction band. 
$D(\ve)=\frac{g|\ve|}{2\pi\gamma^2}$ 
is the density of states of single layer graphene 
with $g=4$ owing to the valley and spin degeneracy. 
$\theta$ is the scattering angle, and $q=2\kf
\sin({\theta}/{2})$ is the scattering wave vector on the circular
two-dimensional Fermi surface as shown in Fig.~\ref{fig:1}(b). The Fermi wave number on each
graphene layer is given as $\kf=\sqrt{{4\pi n_{c}}/{g}}$. 
In the present assumption, since the carrier concentration for each layer
is identical,
the interlayer scattering rate is equivalent,
i.e. ${\tau^{-1}_{12}}={\tau^{-1}_{21}}$. 
Note that the last $\theta$
dependent factor also contains the phase of the wave
function of graphene\cite{Ando2006,Nomura2006}. 
Structural parameter such as the interlayer distance $d$ and
scattering potential due to charged impurities are included in the
screened potentials $W_{ij}$.

Here we briefly explain the derivation of the screened Coulomb potential
$W_{ij}(q,d)$ from
the unscreened one $v_{ij}(q,d)$. The analytical expression of the unscreened Coulomb
potentials of the GDLS can be derived
with the image charge method~\cite{Kumagai1989, Jena2007}. 
For this
system, we need to consider an infinite series of point image charges
arising from two interfaces at $z = 0$ and $d$ shown in Fig.1(a), where two types of
dielectrics are spanned by a graphene layer. 
The resulting unscreened Coulomb potentials are given as
\begin{align}
v_{11}(q, d)=&\frac{4\pi e^2}{q}\frac{\e_2+\e_3\tanh(qd)}{\e_2(\e_1+\e_3)+(\e_2^2+\e_1\e_3)\tanh(qd)},
\\
v_{22}(q, d) =&\frac{4\pi e^2}{q}\frac{\e_2+\e_1\tanh(qd)}{\e_2(\e_1+\e_3)+(\e_2^2+\e_1\e_3)\tanh(qd)},
\\
v_{12}(q, d)=&\frac{4\pi e^2}{q}\frac{\e_2\frac{1}{\cosh(qd)}}{\e_2(\e_1+\e_3)+(\e_2^2+\e_1\e_3)\tanh(qd)}.
\end{align}
Here $v_{11}(q,d)$ and $v_{22}(q,d)$ are the intralayer Coulomb interactions
on the first and second graphene layer, respectively. $v_{12}(q,d)$ is the interlayer Coulomb
interaction. 
These potentials have been used in the context of the superfluid
magnetoexitons \cite{Pikalov2012} and plasmon mode \cite{Badalyan2012,
Profumo2012} of the GDLS. 
The above expressions indicate that the parameter $qd$ ($\approx \kf d$) determines the
screening behavior and the strength of interlayer Coulomb interaction. 
For $qd\gg 1$, the interlayer interaction becomes negligible $v_{12}(q,d)\approx 0$;
that is, the two graphene layers are independent. In contrast, 
for $qd\ll 1$, the two graphene layers are rather strongly bounded through
the interlayer Coulomb interaction. 
The screened Coulomb potentials are given by the following RPA equation, 
\begin{align}
W_{ij}(q, d)\equiv v_{ij}(q, d)+v_{ik}(q, d)\Pi_{kl}(q)W_{lj}(q,d),
\end{align}
where $\Pi_{kl}(q)= \Pi_G(q)\delta_{kl}$. $\Pi_G(q)$ is the static
polarization function of single layer graphene \cite{Ando2006a,
Hwang2007}. By combining Eqs. (2)-(6), we can evaluate the carrier
mobility and scattering rate of the GDLS.

We first investigate the dependence of the mobility 
on the interlayer distance. To focus on the role of $\e_2$, we consider three different dielectrics
as the middle layer, that is, $h$-BN, ${\rm Al_2O_3}$ and ${\rm HfO_2}$. 
Those dielectric constants are $\e_{h-{\rm BN}}=4$, $\e_{\rm
Al_2O_3}=12.53$ and $\e_{\rm HfO_2}=22$, respectively~\cite{Gorbachev2012, Frederisk1991, Desgreniers1991, Kang2000}. 
Other ultrathin dielectrics with high-$\kappa$ can be candidates
for the inner barrier layer in further studies\cite{Wilk2001, Osada2012}. 
Here we assume that the top and bottom dielectrics are air and ${\rm Al_2O_3}$,
respectively: 
$\e_1=\e_{\rm Air}=1$ and $\e_3=\e_{\rm  Al_2O_3}=12.53$.

Figure~\ref{fig:1}(c) shows the dependence of mobility on the interlayer
distance $d$ for three different middle dielectrics $\e_2$. The
mobilities are independent of $\e_2$ at a smaller
interlayer distance ($d\lesssim 0.1$ nm).  
With increasing interlayer distance, 
the effect of the middle layer dielectric becomes prominent. 
In the large $d$ ($\gtrsim 10$ nm) limit, that is, $\kf d\gg 1$, 
the carrier mobilities strongly depend on $\e_2$,
because the interlayer Coulomb interaction becomes negligible.
Thus the carrier mobility can be improved by inserting the higher
dielectrics as the middle layer. 

The strong $\e_2$ dependence for the thicker $d$ region can be 
attributed to the decay of interlayer scattering. 
Figure~\ref{fig:1}(d) shows the $d$-dependence of three components of
the scattering rate
($\tau^{-1}_{11}$,$\tau^{-1}_{22},2\tau^{-1}_{12}$) and the total scattering rate
($\tau^{-1}_{\rm tot}$) for three different dielectrics.
The intralayer components, ${\tau^{-1}_{11}}$ and ${\tau^{-1}_{22}}$,
strongly depend on $\e_2$, whereas the interlayer component ${2}{\tau^{-1}_{12}}$ is
almost independent of $\e_2$ and decays for wider $d$ regions. 

Figure~\ref{fig2}(a) shows the dependence of the carrier mobility on 
$\e_2$ for several fixed interlayer distances. 
We can see that the carrier mobility monotonically increases 
with the $\e_2/\e_3$ ratio when the interlayer distance
becomes wider; see the case of $d=3$ and $15$ nm. 
However, for the thinner case $d=0.7$ nm, i.e., the mobility increases
more quickly when
${\e_2}/{\e_3}\lesssim 1$, however its growth rate slows
at ${\e_2}/{\e_3}\gtrsim 1$. 
Thus, the combination of dielectrics with 
${\e_2}/{\e_3}\lesssim 1$ should be chosen to improve the mobility for
the thinner GDLS. 

The nonmonotonic behavior for the $d = 0.7$ nm case can be 
related to the development of the interlayer scattering rate $\tau^{-1}_{12}$.
Figure~\ref{fig2}(b) shows the total scattering
rate ${\tau^{-1}_{tot}}$ together with its components ${\tau^{-1}_{11}}$,
${\tau^{-1}_{22}}$ and ${2}{\tau^{-1}_{12}}$ as the function of the
middle layer dielectric constant for $d=0.7$ (left panel) and $15$ nm
(right panel).  
Here we can see that the interlayer component ${2}{\tau^{-1}_{12}}$
grows in the higher $\e_2$ region for the case $d = 0.7$ nm owing to the stronger interlayer interaction. 
For this reason, the mobility at $d=0.7$ nm in the lower $\e_2$ region is improved. 
In higher $\e_2$ region, however, the mobility at $d=0.7$ nm becomes lower
than the mobility at $d = 3$ and $15$ nm owing to growth of the interlayer
component ${2}{\tau^{-1}_{12}}$. This interlayer screening effect is
effective at $\kf d \lesssim 1$ and is negligible at $\kf d \gtrsim 1$. 
The enhancement of interlayer screening effect for the region of $\kf d \lesssim 1$ indicates that the carrier transport properties are strongly affected by the interlayer interaction due to the charged impurities. Since such scattering mechanism has not been considered in conventional theory of Coulomb drag, it might be necessary to include the interlayer screening effect for GDLS with small $d$, which is available in recent experiment \cite{Gorbachev2012}. 

So far, we have fixed the dielectric constants for the top and bottom layers
at $\e_1=1$ and $\e_3=12.53$, respectively. Next we focus on
the effect of the dielectric environment from top or bottom layers 
on the carrier mobility at $d=0.7$ nm, which satisfies the criteria of
$\kf d \ll 1$, that is, the interlayer screening is not negligible. 
Figure~\ref{fig2}(c) shows the carrier mobility as a function of
 $\e_2$ for several ratios of the dielectric constants on the top and bottom layer.
This figure clearly shows that the
mobility gradually improve with increasing $\e_1/\e_3$ and that the
interlayer screening effect also works in higher region of $\e_1/\e_3$.  
Therefore, the region of $\e_2 $ which the mobility at $d=0.7$ nm is higher
than that at $d=3$ and $15$ nm, widens with increasing
$\e_1/\e_3$.  
 
In conclusion, we have investigated the dielectric environment effect on 
the charged-impurity-limited carrier mobility of the GDLS on the basis
of the Boltzmann transport
theory. The carrier mobility of GDLS strongly depends on the 
interlayer distance, and can be elevated drastically by using high
$\kappa$ materials as the middle layer. Our result shows that the
mobility can be improved by inserting higher dielectric materials
into thicker GDLS. For thinner GDLS, however, the mobility
can be improved by choosing a combination of dielectrics with
$\e_2/\e_3\lesssim 1$. This condition originates from 
the interlayer screening effect due to strong interlayer
coupling. Our results are intended as guidelines for experiments and
applications of new functional atomically-thin devices.  

This work is supported by a Grant-in-Aid for Scientific Research (KAKENHI)
(Nos. 24710153 and 23310083) from the Japan Society for the Promotion of Science.

\end{document}